\documentclass[a4paper,amsmath,amssymb]{jpconf}
\usepackage{graphicx}
\usepackage{amsmath, amsthm, amssymb}
\usepackage{bm}

\begin{document}
\title{Pattern-recalling processes in quantum Hopfield networks far from saturation}

\author{Jun-ichi Inoue}

\address{Graduate School of Information Science and Technology, Hokkaido University, 
N14-W9, Kita-ku, Sapporo 060-0814, Japan
}

\ead{j$\underline{\,\,\,}$inoue@complex.ist.hokudai.ac.jp}

\begin{abstract}
As a mathematical model of associative memories, 
the Hopfield model was  now well-established 
and a lot of studies 
to reveal the pattern-recalling process have been done 
from various different approaches. As well-known, 
a single neuron is itself an uncertain, noisy unit with a finite unnegligible error in 
the input-output relation. 
To model the situation artificially,  a kind of `heat bath' 
that surrounds neurons is introduced.  The heat bath, which 
is a source of noise, is specified by the `temperature'. 
Several studies concerning 
the pattern-recalling processes of the Hopfield model 
governed  by the Glauber-dynamics at finite temperature 
were already reported. 
However, we might extend the `thermal noise' to 
the quantum-mechanical variant.  In this paper, 
in terms of the stochastic process of quantum-mechanical Markov chain Monte Carlo method (the quantum MCMC), 
we analytically derive macroscopically deterministic equations of order parameters 
such as `overlap' in a quantum-mechanical variant 
of the Hopfield neural networks (let us call {\it quantum Hopfield model} or 
{\it quantum Hopfield networks}). 
For the case in which non-extensive number $p$ of patterns are embedded via  
asymmetric Hebbian connections, 
namely, $p/N \to 0$ for 
the number of neuron $N \to \infty$ (`far from saturation'), 
we evaluate the recalling processes for one of the built-in patterns under the influence of quantum-mechanical noise. 
\end{abstract}
\section{Introduction}
\label{sec:Introduction}
Basic concept of associative memories 
in artificial neural networks was already proposed 
in early 70' s by a Japanese engineer Kaoru Nakano \cite{Nakano1972}. 
Unfortunately, in that time, 
nobody interested in his model, however in 80's, J.J. Hopfield \cite{Hopfield,Mezard}  
pointed out that there exists an energy function 
(Lyapunov function)  
in the so-called {\it associatron} (the Nakano model) and 
the system can be treated as a kind of spin glasses. 
After his study, a lot of researchers who were working 
in the research field of condensed matter physics 
picked the so-called {\it Hopfield model} up for their brand-new `target materials'. 

Among these studies, a remarkable progress 
has been done by  three theoretical physicists Amit, Gutfreund and Sompolinsky \cite{Amit1985} 
who clearly (mathematically) defined the concept of {\it storage capacity} in the 
Hopfield model as a critical point at which 
system undergoes phase transitions from ferromagnetic retrieval to 
spin glass phases by utilizing the replica method. 
They also introduced a noise to prevent the network 
from retrieving one of the built-in patterns as  
a `heat bath' which surrounds the neurons.
They  draw the phase diagram which is composed of three distinct phases, 
namely, ferromagnetic-retrieval, paramagnetic and spin glass phases. 
These phase boundaries are specified by two 
control parameters, namely, 
storage capacity and temperature of the heat bath. 

To evaluate the storage capacity of 
the Hopfield model without energy function 
(for instance, Hopfield model having a non-monotonic input-output 
function \cite{Inoue1996}), 
Shiino and Fukai \cite{Shiino1992}
proposed the so-called 
Self-Consistent Signal-to-Noise Analysis (SCSNA) 
which enables us to 
derive a couple of self-consistent macroscopic 
equations of state by making use of 
the concept of the TAP equations \cite{Mezard}. 

As we mentioned, these theoretical arguments are 
constructed for the case in which 
each neuron is surrounded by a heat bath at finite temperature. 
In this sense, the above studies revealed 
the robustness of the associative memories 
against {\it thermal noises} in the artificial brain. 

However, we might consider 
a different kind of such noises, that is, 
{\it quantum-mechanical noise}. 
As such successful attempts, 
Ma and Gong \cite{Ma}, 
Nishimori and Nonomura \cite{Nishi1996} 
independently 
introduced the quantum-mechanical noise 
to the conventional Hopfield model by 
adding the transverse field to 
the classical Hamiltonian (from now on, we 
shall refer the model as {\it quantum Hopfield model}). 
Especially, Nishimori and Nonomura  \cite{Nishi1996} 
investigated the structure of 
retrieval phase diagrams by using of 
the standard replica method with the assistance of the 
static approximation.  
Therefore, we might say that 
the equilibrium properties of the Hopfield model were now well-understood and 
the methodology to investigate the model was already established. 

On the other hand, theory of the dynamics to evaluate 
the pattern-recalling processes is not yet well-established. 
However,  up to now, several powerful approaches were proposed. 
For instance, Amari and Maginu \cite{Amari} 
pointed out that the relevant macroscopic 
quantities in the synchronous neuro-dynamics 
are the overlap (the direction cosine) and the noise variance. 
They derived the update equations with respect to these quantities. 
After their study, the so-called Amari-Maginu theory was improved by 
taking into account the correlations in 
the noise variances by Okada \cite{Okada}. 

Whereas for the asynchronous dynamics, 
Coolen and his co-authors 
established a general approach, 
so-called {\it dynamical replica theory} \cite{Coolen1994,Coolen1996}. 
They utilized two assumptions, 
namely, equipartitioning in the sub-shells and 
self-averaging of the intrinsic noise distribution to 
derive the deterministic flow 
equations for relevant macroscopic quantities. 
However, there is no such theoretical framework so far to 
investigate the pattern-recalling process 
of the quantum Hopfield model systematically. 

In this paper, we propose such a candidate 
of dynamical theory to deal with 
the pattern-recalling processes in 
the quantum Hopfield model. 
We shall consider the stochastic process 
of quantum Monte Carlo method 
which is applied for the quantum Hopfield model 
and investigate the quantum neuro-dynamics 
through the differential equations 
with respect to the macroscopic quantities 
such as the overlap. 

This paper is organized as follows. 
In the next section 2, 
we explain the basics of 
the conventional Hopfield model and its generic properties. 
Then, we categorize the model into two distinct classes 
according to the origin of noises in artificial brain.   
The quantum Hopfield model is 
clearly defined. 
In section 3, 
we explain the quantum Monte Carlo method 
based on the Suzuki-Trotter decomposition \cite{Suzuki2}
and consider the stochastic process 
in order to investigate the 
pattern-recalling dynamics of the quantum 
Hopfield model. 
In section 4,  we 
derive the macroscopic deterministic flow of 
the overlap between the neuronal state and 
one of the built-in patterns 
from the microscopic master equation 
which describes the stochastic 
processes in the quantum Monte Calro method 
for the quantum Hopfield model \cite{Inoue2010}. 
The general solution of the dynamics is 
obtained under the so-called static approximation. 
In the next section 5, 
we apply our general solution to 
a special case, namely, 
sequential recalling the built-in two patterns 
via asymmetric Hebb connections. 
The effect of quantum-mechanical noise is 
compared with that of the conventional thermal noise. 
The last section is summary. 
\section{The Hopfield model}
\label{sec:Hopfield}
In this section, we briefly 
explain the basics of the conventional Hopfield model. 
Then, we shall divide the model into 
two classes, namely, 
the Hopfield model 
put in thermal noises (the model 
is referred to as {\it classical systems}) and 
the same model in the quantum-mechanical noise 
(the model is referred to as {\it quantum systems}). 
In following, we define each of the models explicitly. 
\subsection{The classical system}
Let us consider the network having $N$-neurons. 
Each neuron $S_{i}$ takes 
two states, namely, 
$S_{i}=+1$ (fire) and 
$S_{i}=-1$ (stationary). 
Neuronal states 
are given by the set of variables $S_{i}$, that is, 
 $\bm{S}=(S_{1},\cdots,S_{N}), \,
S_{i} \in \{+1,-1\}$. 
Each neuron is located on a complete graph, 
namely, graph topology of the network is `fully-connected'.  
The synaptic connection between 
arbitrary two neurons, say, 
$S_{i}$ and $S_{j}$ is defined by the following 
Hebb rule:  
\begin{equation}
J_{ij}   = 
\frac{1}{N} \sum_{\mu,\nu}
\xi_{i}^{\mu}A_{\mu\nu}\xi_{j}^{\nu}
\end{equation}
where 
$\bm{\xi}^{\mu}=(\xi_{1},\cdots,\xi_{N}), 
\xi_{i}^{\mu} \in \{+1,-1\}$ denote 
the embedded patterns 
 and each of them is specified 
 by a label $\mu=1,\cdots,P$. 
 $A_{\mu\nu}$ denotes 
$(P \times P)$-size matrix and 
$P$ stands for the number of 
built-in patterns. We should keep in mind that 
there exists an energy function (a Lyapunov function) 
in the system if the matrix 
$A_{\mu\nu}$ is symmetric. 

Then, the output of the neuron $i$, that is, 
$S_{i}$ is determined by the sign of the local field 
$h_{i}$ as 
\begin{equation}
h_{i} =   
\sum_{\mu,\nu=1}^{p}
\xi_{i}^{\mu}
A_{\mu\nu}m^{\nu} + 
\frac{1}{N}
\sum_{\mu^{'},\nu^{'}=p+1}^{P}
\xi_{i}^{\mu^{'}}
A_{\mu^{'}\nu^{'}}
\sum_{j}\xi_{j}^{\nu^{'}}
S_{j}
\label{eq:local}
\end{equation}
where 
$A_{\mu\nu}$ and $A_{\mu^{'} \nu^{'}}$ are elements 
of $p \times p, (P-p) \times (P-p)$-size matrices, 
respectively. 
We also defined the overlap (the direction cosine) between 
the state of neurons $\bm{S}$ and one of the built-in patterns 
$\bm{\xi}^{\nu}$ by 
\begin{eqnarray}
m^{\nu} & \equiv & 
\frac{1}{N}\,(\bm{S} \cdot \bm{\xi}^{\nu}) = 
\frac{1}{N}\sum_{i}\xi_{i}^{\nu}S_{i}.
\end{eqnarray}
Here we should notice that 
the Hamiltonian of the system is given by 
$-\sum_{i}h_{i}S_{i}$. 
The first term appearing in the left hand side of equation 
(\ref{eq:local}) is a contribution 
from $p \sim {\cal O}(1)$ what we call `condensed patterns', 
whereas the second term stands for 
the so-called `cross-talk noise'. 
In this paper, we shall concentrate ourselves to 
the case in which 
the second term is negligibly small in 
comparison with the first term, 
namely, the case of 
$P=p \sim \mathcal{O}(1)$. 
In this sense, we can say that 
the network is `far from its saturation'. 
\subsection{The quantum system}
To extend the classical system to the quantum-mechanical variant, 
we rewrite the local field $h_{i}$ as follows. 
\begin{equation}
\bm{\phi}_{i} = 
\sum_{\mu,\nu=1}^{p}
\xi_{i}^{\mu}
A_{\mu\nu}
\left(
\frac{1}{N}\sum_{i}\xi_{i}^{\nu}
\bm{\sigma}_{i}^{z}
\right)
\end{equation}
where $\bm{\sigma}^{z}_{i}$ ($i=1,\cdots,N$)   
stands for the $z$-component of the Pauli matrix. 
Thus, the Hamiltonian 
$\bm{H}_{0} \equiv 
-\sum_{i}\bm{\phi}_{i}\bm{\sigma}_{i}^{z}$ is 
a diagonalized $(2^{N} \times 2^{N})$-size 
matrix and the lowest eigenvalue is 
identical to the ground state of the classical 
Hamiltonian $-\sum_{i}\phi_{i}S_{i}$ 
($S_{i}$ is an eigenvalue of the matrix $\bm{\sigma}_{i}^{z}$). 

Then, we introduce quantum-mechanical noise into 
the Hopfield neural network by adding 
the transverse field to the Hamiltonian as follows. 
\begin{equation}
\bm{H} = 
\bm{H}_{0}-
\Gamma \sum_{i=1}^{N}\bm{\sigma}_{i}^{x}
\label{eq:hami1}
\end{equation}
where $\bm{\sigma}_{i}^{x}$ is 
the $x$-component of the Pauli matrix and 
transitions between 
eigenvectors of the classical Hamiltonian $\bm{H}_{0}$ are 
induced due to the off-diagonal elements of 
the matrix $\bm{H}$ for 
$\Gamma \neq 0$. 
In this paper, we mainly consider 
the system described by 
(\ref{eq:hami1}). 
\section{Quantum Monte Carlo method}
The dynamics of the quantum 
model (\ref{eq:hami1}) follows 
Schr$\ddot{\rm o}$dinger equation. 
Thus, we should solve it or 
investigate the 
time dependence of the state 
$|\psi (t) \rangle$ 
by using the time-evolutionary operator $\mbox{\rm e}^{-i \bm{H}\Delta t /\hbar}$ 
defined for infinitesimal time $\Delta t$ as
\begin{eqnarray}
 |\psi (t +\Delta t) \rangle & = &  
\mbox{\rm e}^{-i\bm{H}\Delta t/\hbar} |\psi (t) \rangle. 
\end{eqnarray}
However, 
even if we carry it out numerically, 
it is very hard for us to do it with reliable precision 
because $(2^{N} \times 2^{N})$-size 
Hamilton matrix becomes huge for 
the number of neurons $N \gg 1$ as in a realistic brain. 
Hence, here we use the quantum Monte Carlo method 
to simulate the quantum system in our personal computer and 
consider the stochastic processes of Glauber-type to 
discuss the pattern-recalling dynamics of the quantum Hopfield model. 
\subsection{The Suzuki-Trotter decomposition}
The difficulty to carry out algebraic calculations in 
the model system  
is due to the non-commutation operators 
appearing in 
the Hamiltonian (\ref{eq:hami1}), 
namely, 
$\bm{H}_{0}, \bm{H}_{1} \equiv -\Gamma \sum_{i}\bm{\sigma}_{i}^{x}$. 
Thus, we use 
the following 
Suzuki-Trotter decomposition \cite{Suzuki2} 
in order to deal with the system as 
a classical spin system. 
\begin{equation}
\mbox{\rm tr}\, 
\mbox{\rm e}^{
\beta (\bm{H}_{0}+\bm{H}_{1})
} =  
\lim_{M \to \infty} 
\mbox{\rm tr}
\left(
{\exp}
\left(
\frac{\beta \bm{H}_{0}}
{M}
\right)
{\exp}
\left(
\frac{\beta \bm{H}_{1}}
{M}
\right)
\right)^{M}
\label{eq:ST}
\end{equation}
where 
$\beta$ denotes the `inverse temperature' 
and  
$M$ is the number of 
the Trotter slices, for which the limit 
$M \to \infty$ should be taken. 
Thus, one can deal with 
$d$-dimensional quantum 
system as the corresponding 
($d+1$)-dimensional 
classical system. 
\section{Derivation of the deterministic flows}
In the previous section, we mentioned that 
we should simulate the quantum 
Hopfield model 
by means of the quantum Monte Carlo method 
to reveal the quantum neuro-dynamics through the time-dependence of 
the macroscopic quantities such as the overlap. 
However, 
in general, it is also very difficult to 
simulate the quantum-mechanical properties at the ground 
state by a personal computer even for finite size systems ($N,M < \infty$). 

With this fact in mind, 
in this section, we attempt to 
derive the macroscopic 
flow equations 
from the microscopic master equation 
for the classical system regarded as 
the quantum 
system in terms of the Suzuki-Trotter decomposition. 
This approach is efficiently possible because 
the Hopfield model is a fully-connected mean-field model 
such as the Sherrington-Kirkpatrick model \cite{SK} for spin glasses 
and its equilibrium properties are 
completely determined by several order parameters. 
\subsection{The master equation}
After the Suzuki-Trotter decomposition (\ref{eq:ST}), we obtain 
the local field for the neuron $i$ located on the 
$k$-th Trotter slice as follows. 
\begin{eqnarray}
\beta \phi_{i}
(\bm{\sigma}_{k}: \sigma_{i}(k \pm 1)) & = & 
\frac{\beta}{M}
\sum_{\mu,\nu}
\xi_{i}^{\nu}
A_{\mu\nu}
\left\{
\frac{1}{N}\sum_{j}
\xi_{j}^{\nu}
\sigma_{j}(k)
\right\} + \frac{B}{2}
\left\{
\sigma_{i} (k-1) + \sigma_{i}(k+1)
\right\}
\end{eqnarray}
where 
parameter $B$ is related to the amplitude of the 
transverse field (the strength of the quantum-mechanical 
noise) $\Gamma$ by 
\begin{equation}
B = 
\frac{1}{2} \log \coth 
\left(
\frac{\beta \Gamma}{M}
\right).
\end{equation}
In the classical limit 
$\Gamma \to 0$, 
the parameter
$B$ goes to infinity. 
For the symmetric matrix $A_{\mu\nu}$, 
the Hamiltonian (scaled by $\beta$) of 
the system is given by 
$-\sum_{i}\beta \phi_{i}(\mbox{\boldmath $\sigma$}_{k} : \sigma (k \pm 1))\sigma_{i}(k)$.

Then, the transition probability 
which specifies the Glauber dynamics 
of the system is given by 
$w_{i}(\mbox{\boldmath $\sigma$}_{k}) =  
(1/2)[
1-\sigma_{i}(k) \tanh (\beta \phi_{i}(\mbox{\boldmath $\sigma$}_{k} : \sigma (k \pm 1)))]$. 
More explicitly, 
$w_{i}(\mbox{\boldmath $\sigma$}_{k})$ denotes 
the probability that an arbitrary neuron 
$\sigma_{i}(k)$ changes its state as $\sigma_{i}(k) \to -\sigma_{i}(k)$ within the time unit. 
Therefore, the probability that the neuron 
$\sigma_{i}(k)$ takes $+1$ is obtained by setting 
$\sigma_{i}(k)=-1$ in the above $w_{i}(\mbox{\boldmath $\sigma$}_{k})$ and 
we immediately find 
$\sigma_{i}(k)=\sigma_{i}(k-1)=\sigma_{i}(k+1)$ with 
probability $1$ in the limit of $B \to \infty$ which implies 
the classical limit $\Gamma \to 0$. 

Hence, the probability that a microscopic state 
including the $M$-Trotter slices
$\{ \bm{\sigma}_{k}\} \equiv 
(\bm{\sigma}_{1},\cdots,\bm{\sigma}_{M}), 
\bm{\sigma}_{k} \equiv (\sigma_{1}(k),\cdots,\sigma_{N}(k))$
obeys the following 
master equation: 
\begin{eqnarray}
\frac{dp_{t}(\{\mbox{\boldmath $\sigma$}_{k}\})}
{dt} & = & 
\sum_{k=1}^{M}\sum_{i=1}^{N}[
p_{t}(F_{i} ^{(k)}(\mbox{\boldmath $\sigma$}_{k}))
w_{i}(F_{i}^{(k)}(\mbox{\boldmath $\sigma$}_{k})) -  
p_{t}(\mbox{\boldmath $\sigma$}_{k})
w_{i}(\mbox{\boldmath $\sigma$}_{k})]
\label{eq:master}
\end{eqnarray}
where $F_{i}^{(k)}(\cdot)$ 
denotes a single spin flip operator 
for neuron $i$ on the Trotter slice $k$ 
as $\sigma_{i}(k) \to -\sigma_{i}(k)$. 
When we pick up the overlap between 
neuronal state $\bm{\sigma}_{k}$ and 
one of the built-in patterns $\bm{\xi}^{\nu}$, 
namely, 
\begin{eqnarray}
m_{k}  & \equiv  & \frac{1}{N} \, 
(\bm{\sigma}_{k} \cdot 
\bm{\xi}^{\nu}) = \frac{1}{N} \sum_{i} \xi_{i}^{\nu} \sigma_{i}(k)
\end{eqnarray}
as a relevant macroscopic quantity, 
the joint distribution of 
the set of the overlaps $\{m_{1},\cdots, m_{M}\}$ at time 
$t$ is written in terms of 
the probability for realizations of 
microscopic states 
$p_{t}(\{\mbox{\boldmath $\sigma$}_{k}\})$ 
at the same time $t$ as  
\begin{equation}
P_{t}(m_{1}^{\nu},\cdots, m_{M}^{\nu})   =  
\sum_{\{\mbox{\boldmath $\sigma$}_{k}\}} 
p_{t}(\{\mbox{\boldmath $\sigma$}_{k}\}) 
\prod_{k=1}^{M}
\delta (m_{k}^{\nu}-m_{k}^{\nu}(\mbox{\boldmath $\sigma$}_{k}))
\label{eq:Pt0}
\end{equation}
where we defined the sums by 
\begin{eqnarray}
\sum_{\{\bm{\sigma}_{k}\}} (\cdots) & \equiv & 
\sum_{\bm{\sigma}_{1}} \cdots 
\sum_{\bm{\sigma}_{M}} (\cdots), \,\,\,\,\,
\sum_{\bm{\sigma}_{k}} (\cdots) \equiv  
\sum_{\sigma_{1}(k)=\pm 1} \cdots 
\sum_{\sigma_{N}(k)=\pm 1} (\cdots). 
\end{eqnarray}
Taking the derivative of 
equation (\ref{eq:Pt0}) 
with respect to $t$ and 
substituting 
(\ref{eq:master}) into the result, we have 
the following differential equations for 
the joint distribution 
\begin{eqnarray}
\frac{dP_{t}(m_{1}^{\nu},\cdots,m_{M}^{\nu})}{dt}  & = &  
\sum_{k}\frac{\partial}{\partial m_{k}^{\nu}}
\{
m_{k}^{\nu}P_{t}(m_{1}^{\nu},\cdots,m_{k}^{\nu},\cdots,m_{M}^{\nu})
\}  \nonumber \\
\mbox{} & -  & \sum_{k}  
\frac{\partial}{\partial m_{k}^{\nu}}
{\Biggr \{}
P_{t}(m_{1}^{\nu},\cdots,m_{k}^{\nu},\cdots,m_{M}^{\nu}) \int_{-\infty}^{\infty}
D[\xi^{\nu}] d\xi^{\nu}  \nonumber \\
\mbox{} & \times & 
\frac{
\sum_{\{\mbox{\boldmath $\sigma$}_{k}\}}
p_{t}(\{\mbox{\boldmath $\sigma$}_{k}\})
\xi^{\nu} \tanh[\beta \phi (k)]
\prod_{k,i} \delta (m_{k}^{\nu}-m_{k}^{\nu}(\mbox{\boldmath $\sigma$}_{k})) 
}
{
\sum_{\{\mbox{\boldmath $\sigma$}_{k}\}}
p_{t}(\{\mbox{\boldmath $\sigma$}_{k}\})
\prod_{k} \delta (m_{k}^{\nu}-m_{k}^{\nu}(\mbox{\boldmath $\sigma$}_{k}))}
{\Biggr \}} \nonumber \\
\mbox{} & \times & \delta (\sigma (k +1)-\sigma_{i}(k + 1))  \delta (\sigma (k -1)-\sigma_{i}(k - 1)) 
\label{eq:dPmkdt}
\end{eqnarray}
where we introduced several notations 
\begin{equation}
D[\xi^{\nu}] \equiv  
\frac{1}{N}\sum_{i}\delta (\xi^{\nu}-\xi_{i}^{\nu})
\label{eq:defDxi}
\end{equation}
\begin{equation}
\beta \phi (k) \equiv 
\frac{\beta \sum_{\mu\nu}\xi^{\mu}A_{\mu\nu}}{M}
m_{k}^{\nu}+\frac{B}{2} \sigma (k-1) + \frac{B}{2}\sigma (k+1) 
\label{eq:local2}
\end{equation}
for simplicity. 

Here we should notice that 
if the local field $\beta \phi(k)$ is 
independent of the microscopic 
variable $\{\bm{\sigma}_{k}\}$, 
one can get around the complicated expectation of 
the quantity $\tanh[\beta \phi(k)]$ 
over the time-dependent Gibbs measurement 
which is defined in the sub-shell: $\prod_{k} \delta (m_{k}^{\nu}-m_{k}^{\nu}(\mbox{\boldmath $\sigma$}_{k}))$. 
As the result,  
only procedure we should carry out to 
get the deterministic flow is to calculate the 
data average (the average over the built-in patterns). 
However, 
unfortunately, 
we clearly find from equation 
(\ref{eq:local2}) that 
the local field depends on the 
$\{\bm{\sigma}_{k}\}$. 
To overcome the difficulty and 
to carry out the calculation, 
we assume that the probability $p_{t}(\{\bm{\sigma}_{k}\})$ of 
realizations for microscopic states 
during the dynamics is 
independent of $t$, 
namely, 
\begin{eqnarray}
p_{t}(\{\bm{\sigma}_{k}\}) & = & p(\{\bm{\sigma}_{k}\}).
\end{eqnarray}
Then, our average over the time-dependent Gibbs measurement 
in the sub-shell is rewritten as 
\begin{eqnarray}
&& \frac{
\sum_{\{\mbox{\boldmath $\sigma$}_{k}\}}
p_{t}(\{\mbox{\boldmath $\sigma$}_{k}\})
\xi^{\nu} \tanh[\beta \phi (k)]
\prod_{k,i} \delta (m_{k}^{\nu}-m_{k}^{\nu}(\mbox{\boldmath $\sigma$}_{k})) 
}
{
\sum_{\{\mbox{\boldmath $\sigma$}_{k}\}}
p_{t}(\{\mbox{\boldmath $\sigma$}_{k}\})
\prod_{k} \delta (m_{k}^{\nu}-m_{k}^{\nu}(\mbox{\boldmath $\sigma$}_{k}))} \nonumber \\
\mbox{} & \times & 
 \delta (\sigma (k +1)-\sigma_{i}(k + 1)) \delta (\sigma (k -1)-\sigma_{i}(k - 1))\nonumber \\
\mbox{} & \equiv & 
\langle 
\xi^{\nu} \tanh[\beta \phi (k)]
\prod_{i} \delta (\sigma (k + 1)-\sigma_{i}(k + 1)) \delta (\sigma (k -1)-\sigma_{i}(k - 1))
\rangle_{*}
\label{eq:subshell}
\end{eqnarray}
where $\langle \cdots \rangle_{*}$ stands for 
the average in the sub-shell defined by 
$m_{k}^{\nu}=m_{k}^{\nu}(\bm{\sigma}_{k})\, (\forall_{k})$: 
\begin{equation}
\langle \cdots \rangle_{*} \equiv  
\frac{
\sum_{\{\mbox{\boldmath $\sigma$}_{k}\}}
p(\{\mbox{\boldmath $\sigma$}_{k}\})
(\cdots)\prod_{k} \delta (m_{k}^{\nu}-m_{k}^{\nu}(\mbox{\boldmath $\sigma$}_{k}))
}
{
\sum_{\{\mbox{\boldmath $\sigma$}_{k}\}}
p(\{\mbox{\boldmath $\sigma$}_{k}\})
\prod_{k} \delta (m_{k}^{\nu}-m_{k}^{\nu}(\mbox{\boldmath $\sigma$}_{k}))}
\label{eq:gibbs}
\end{equation}
If we notice that 
the Gibbs measurement 
in the sub-shell is rewritten as  
\begin{eqnarray}
\sum_{\{\mbox{\boldmath $\sigma$}_{k}\}}
p(\{\mbox{\boldmath $\sigma$}_{k}\}) 
\prod_{k} \delta (m_{k}^{\nu}-m_{k}^{\nu}(\mbox{\boldmath $\sigma$}_{k}))   & = & 
\mbox{\rm tr}_{\{\sigma\}}
{\exp}
\left[
\beta \sum_{l=1}^{M}\phi (l) \sigma (l)
\right]
\end{eqnarray}
($\mbox{\rm tr}_{\{\sigma\}}(\cdots) \equiv 
\prod_{k}
\sum_{\mbox{\boldmath $\sigma$}_{k}}(\cdots)$), 
and the quantity 
\begin{equation}
\tanh
\left[
\beta \phi (k) 
\right] = 
\frac{\sum_{\sigma (k)=\pm 1}
\sigma (k) \,{\exp}[\beta \phi (k)\sigma(k)]}
{\sum_{\sigma (k)=\pm 1}
{\exp}[\beta \phi (k)\sigma(k)]} 
\end{equation}
is independent of $\sigma(k)$, 
the average appearing in  
(\ref{eq:subshell}) leads to 
\begin{eqnarray}
\langle 
\xi^{\nu} \tanh[\beta \phi (k)]
\prod_{i} \delta (\sigma (k \pm 1)-\sigma_{i}(k \pm 1))
\rangle_{*} & = & 
\frac{\mbox{\rm tr}_{\{\sigma\}}
\xi^{\nu}
\{
\frac{1}{M}
\sum_{l=1}^{M}
\sigma (l)
\}
 \exp [
\beta \phi (k)\sigma (k)]
}
{{\rm tr}_{\{\sigma\}}
\exp  [
\beta \phi (k) \sigma (k)
]} \nonumber \\
\mbox{} & \equiv  &  \xi^{\nu} \langle \sigma  \rangle_{path}^{(\xi^{\nu})} 
\end{eqnarray}
in the limit of $M \to \infty$. 
This is nothing but a path integral for the effective single neuron problem 
in which 
the neuron 
updates its state along the imaginary time axis: 
$\mbox{\rm tr}_{\{\sigma\}} (\cdots) \equiv 
\sum_{\sigma (1)=\pm 1}\cdots 
\sum_{\sigma (M)=\pm 1}(\cdots)$ 
with weights ${\exp}[\beta \phi (k) \sigma (k)],\,
(k=1,\cdots,M)$.   

Then, the differential equation 
(\ref{eq:dPmkdt}) leads to 
\begin{eqnarray}
\frac{dP_{t}(m_{1}^{\nu},\cdots,m_{M}^{\nu})}
{dt}   & = & 
\sum_{k}
\frac{\partial}{\partial m_{k}^{\nu}}
\{
m_{k}^{\nu}P_{t}(m_{1}^{\nu},\cdots,m_{k}^{\nu},\cdots,m_{M}^{\nu})
\}  \nonumber \\
\mbox{} & - & 
\sum_{k}  
\frac{\partial}{\partial m_{k}^{\nu}}
{\Biggr \{}
P_{t}(m_{1}^{\nu},\cdots,m_{k}^{\nu},\cdots,m_{M}^{\nu}) 
\int_{-\infty}^{\infty}
D[\xi^{\nu}] d\xi^{\nu}  \xi^{\nu} \langle \sigma \rangle_{path}^{(\xi^{\nu})} 
{\Biggr \}}.
\label{eq:dPt}
\end{eqnarray}
In order to derive the compact form of the differential equations 
with respect to the overlaps, we substitute 
$P_{t}(m_{1}^{\nu},\cdots,m_{M}^{\nu}) = 
\prod_{k=1}^{M}
\delta (m_{k}^{\nu}-m_{k}^{\nu}(t))$
into the above 
(\ref{eq:dPt}) and 
multiplying 
$m_{l}^{\nu}$ by both sides of the equation and 
carrying out the integral 
with respect to 
$dm_{1}^{\nu}\cdots dm_{M}^{\nu}$ by part, 
we have for  $l=1,\cdots, M$ as 
\begin{eqnarray}
\frac{d m_{l}^{\nu}}{dt} & = & 
-m_{l}^{\nu}+
\int_{-\infty}^{\infty} D[\xi^{\nu}]
d\xi^{\nu}\xi^{\nu}
\langle \sigma \rangle_{path}^{(\xi^{\nu})}. 
\label{eq:dmdtQ}
\end{eqnarray}
Here we should notice that 
the path integral 
$\xi^{\nu} \langle \sigma  \rangle_{path}^{(\xi^{\nu})}$ 
depends on the embedded 
patterns $\bm{\xi}^{\nu}$. 
In the next subsection, we carry out the 
quenched average explicitly 
under the so-called static approximation. 
\subsection{Static approximation}
In order to obtain the final form of the deterministic flow, 
we assume that macroscopic quantities such as 
the overlap are independent of 
the Trotter slices $k$ during the dynamics. 
Namely, we must use the so-called static approximation: 
\begin{eqnarray}
m_{k}^{\nu} & =  & m^{\nu}\,(\forall_{k}). 
\end{eqnarray}
Under the static approximation, 
let us use the following 
inverse process of the 
Suzuki-Trotter decomposition (\ref{eq:ST}): 
\begin{equation}
\lim_{M \to \infty} 
Z_{M} = 
\mbox{\rm tr}\,{\exp}
\left[
\beta \sum_{\mu\nu}\xi^{\mu}A_{\mu\nu} m^{\nu} \sigma_{z} + \beta \Gamma \sigma_{x}
\right] 
\end{equation}
\begin{eqnarray}
Z_{M} & \equiv &  
\mbox{\rm tr}_{\{\sigma\}}
{\exp}
{\Biggr [}
\frac{\beta \sum_{\mu\nu}\xi^{\mu}A_{\mu\nu} m^{\nu}}{M}
\sum_{k}\sigma (k) + 
B \sum_{k}\sigma (k)\sigma (k+1)
{\Biggr ]}
\end{eqnarray}
In our previous study \cite{Inoue2010}, 
we numerically checked the validity of static approximation 
by computer simulations and found that 
the approximation is successfully valid for 
the pure-ferromagnetic system but 
it is deviated from the good approximation for the disordered systems. 
The validity of the static approximation 
was recently argued by Takahashi and Matsuda \cite{Takahashi} 
from the different perspective. 

Then, one can calculate the path integral immediately as 
\begin{eqnarray}
 \langle \sigma \rangle_{path}^{(\xi^{\nu})} & = &     
\frac{\sum_{\mu\nu}\xi^{\mu}A_{\mu\nu} m^{\nu}}
{\sqrt{(\sum_{\mu\nu}\xi^{\mu}A_{\mu\nu} m^{\nu})^{2}+\Gamma^{2}}} 
\tanh 
\beta 
\sqrt{
\left(
\sum_{\mu\nu}\xi^{\mu}A_{\mu\nu} m^{\nu}
\right)^{2}+\Gamma^{2}}. 
\end{eqnarray}
Inserting this result into (\ref{eq:dmdtQ}), 
we obtain 
\begin{eqnarray}
\frac{dm^{\nu}}{dt} & = & 
-m^{\nu}+ \mathbb{E}_{\bm{\xi}}
{\Biggr [}
\frac{\xi^{\nu}\sum_{\mu\nu}\xi^{\mu}A_{\mu\nu} m^{\nu}}
{\sqrt{(\sum_{\mu\nu}\xi^{\mu}A_{\mu\nu} m^{\nu})^{2}+\Gamma^{2}}}  
\tanh 
\beta 
\sqrt{
\left(
\sum_{\mu\nu}\xi^{\mu}A_{\mu\nu} m^{\nu}
\right)^{2}+\Gamma^{2}}
{\Biggr ]}
\label{eq:result}
\end{eqnarray}
where we should bear in mind that 
the empirical distribution $D[\xi^{\nu}]$ in (\ref{eq:dmdtQ})
was replaced 
by the built-in pattern distribution $\mathcal{P}(\xi^{\nu})$ as
\begin{equation}
\lim_{N \to \infty}
\frac{1}{N}
\sum_{i}\delta (\xi_{i}^{\nu}-\xi^{\nu}) = 
\mathcal{P}(\xi^{\nu})
\end{equation}
in the limit of 
$N \to \infty$ 
and the average is now carried out explicitly as 
\begin{eqnarray}
\int D[\xi^{\nu}]d\xi^{\nu}(\cdots) =
\int \mathcal{P}(\xi^{\nu}) d\xi^{\nu}(\cdots) 
\equiv 
\mathbb{E}_{\bm{\xi}}[\cdots]. 
\end{eqnarray} 
Equation (\ref{eq:result}) 
is a general solution for the problem in this paper. 
\subsection{The classical and zero-temperature limits}
It is easy for us to take the classical limit $\Gamma \to 0$ 
in the result (\ref{eq:result}). 
Actually, we have immediately 
\begin{equation}
\frac{dm^{\nu}}{dt}  =  
-m^{\nu}+ \mathbb{E}_{\bm{\xi}}
\left[
\xi^{\nu}
\tanh 
\left(
\beta 
\sum_{\mu\nu}\xi^{\mu}A_{\mu\nu} m^{\nu}
\right)
\right].
\end{equation}
The above equation is 
identical to the result by 
Coolen and Ruijgrok \cite{Coolen1988} 
who considered the retrieval process of 
the conventional Hopfield model 
under thermal noise. 

We can also take the zero-temperature limit 
$\beta \to \infty$ in (\ref{eq:result}) as 
\begin{equation}
\frac{dm^{\nu}}{dt}  =  
-m^{\nu}+ \mathbb{E}_{\bm{\xi}}
\left[
\frac{\xi^{\nu}\sum_{\mu\nu}\xi^{\mu}A_{\mu\nu} m^{\nu}}
{\sqrt{(\sum_{\mu\nu}\xi^{\mu}A_{\mu\nu} m^{\nu})^{2}+\Gamma^{2}}}
\right]. 
\end{equation}
Thus, 
the equation (\ref{eq:result}) including 
the above two limiting cases is our general solution 
for the neuro-dynamics of 
the quantum Hopfield model 
in which 
$\mathcal{O}(1)$ patterns are embedded. 
Thus, we can discuss any kind of 
situations for such pattern-recalling processes 
and the solution is always derived from  (\ref{eq:result}) explicitly. 
\section{Limit cycle solution for asymmetric connections}
In this section, we discuss a special case of 
the general solution (\ref{eq:result}). 
Namely, 
we investigate the 
pattern-recalling processes of the quantum 
Hopfield model with asymmetric 
connections $\bm{A} \equiv \{A_{\mu\nu}\}$.
\subsection{Result for two-patterns }
\begin{figure}[ht]
\begin{center}
\includegraphics[width=9.5cm]{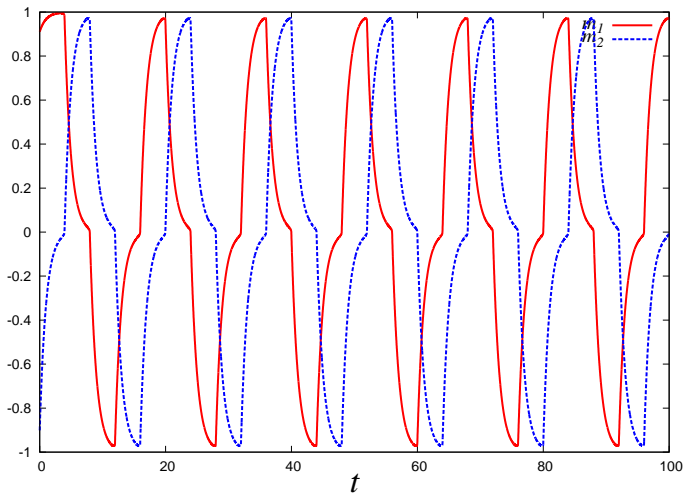}
\end{center}
\caption{\footnotesize 
Time evolutions of  
$m_{1}$ and $m_{2}$ for the case of 
$\Gamma=0.01$. 
 }
\label{fig:fg4}
\end{figure}

Let us consider the case in 
which just only two patterns are embedded via the following matrix: 
\begin{equation}
A=
\left(
\begin{array}{cc}
1 & -1 \\
1 & 1
\end{array}
\right)
\end{equation}
Then, 
from the general solution (\ref{eq:result}), 
the differential equations with respect to 
the two overlaps $m_{1}$ and $m_{2}$ are 
written as 
\begin{eqnarray*}
\frac{dm_{1}}{dt} & = & 
-m_{1}+ 
\frac{m_{1}}{\sqrt{(2m_{1})^{2}+\Gamma^{2}}} - 
\frac{m_{2}}
{\sqrt{(2m_{2})^{2}+\Gamma^{2}}} \\
\frac{dm_{2}}{dt} & = & 
 -m_{2}+ 
\frac{m_{1}}{\sqrt{(2m_{1})^{2}+\Gamma^{2}}} + 
\frac{m_{2}}{\sqrt{(2m_{2})^{2}+\Gamma^{2}}}. 
\end{eqnarray*}
In Figure \ref{fig:fg4}, 
we show the time evolutions of  
the overlaps $m_{1}$ and $m_{2}$ for the case of the amplitude $\Gamma=0.01$. 
From this figure, we clearly find that 
the neuronal state evolves as 
$A \to B \to \overline{A} \to \overline{B} \to A \to B \to \cdots$
($\overline{A},\overline{B}$ denote 
the `mirror images' of 
$A$ and $B$, respectively), 
namely, the network behaves as 
a limit cycle. 

\begin{figure}[ht]
\begin{center}
\mbox{}\hspace{-0.2cm}
\includegraphics[width=8cm]{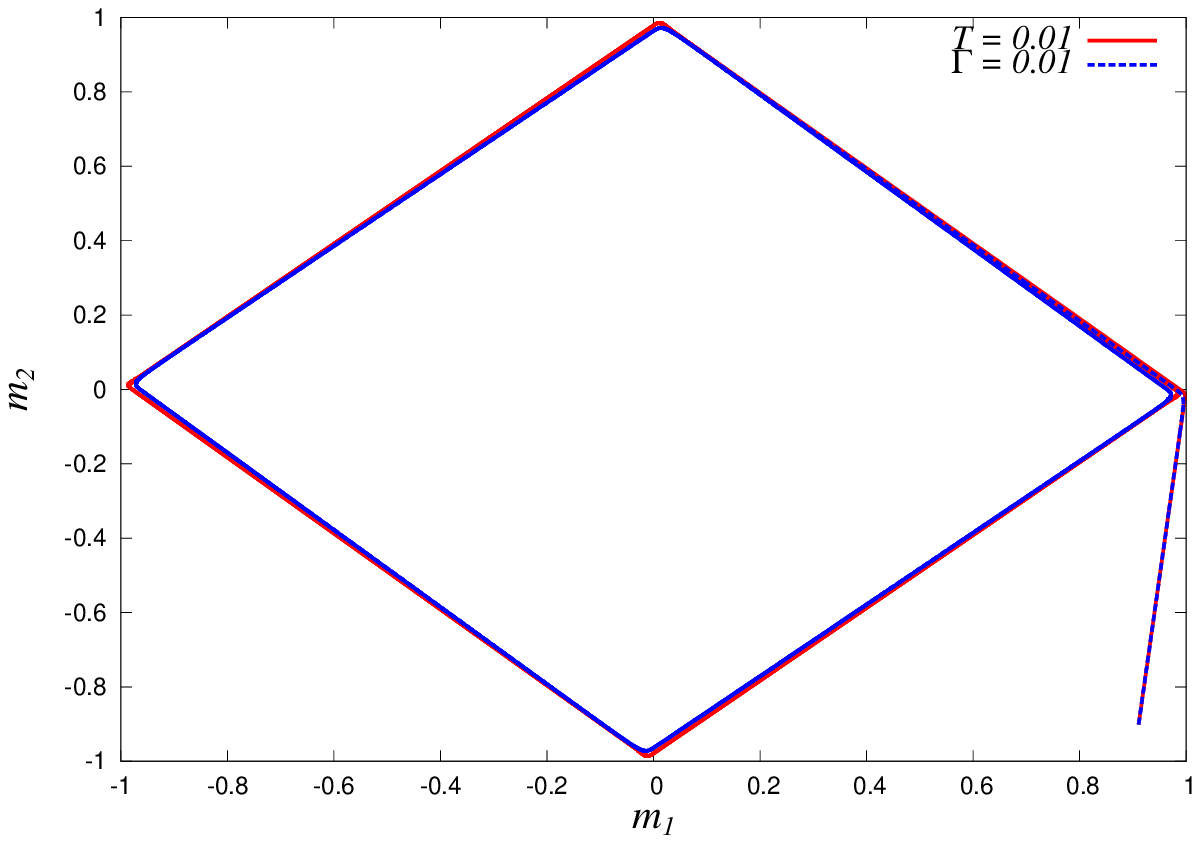} \hspace{-0.3cm}
\includegraphics[width=8cm]{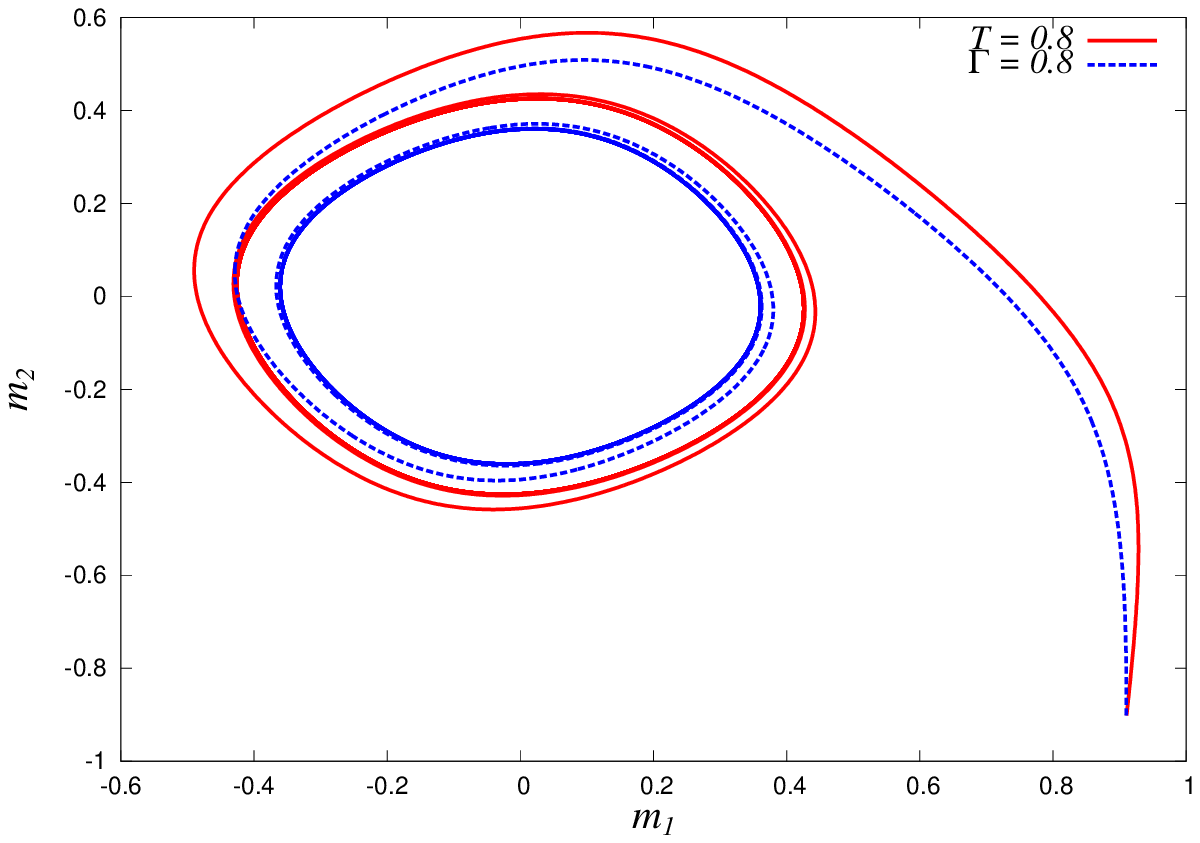}
\end{center}
\caption{\footnotesize 
Trajectories $m_{1}$-$m_{2}$ for 
$(T, \Gamma)=(0, 0.01), (0.01,0)$ (left panel), 
$(T, \Gamma)=(0,0.8),(0.8,0)$ (right panel). }
\label{fig:fg5}
\end{figure}

To compare the effects of  
thermal and quantum noises on the 
pattern-recalling processes, 
we plot the trajectories $m_{1}$-$m_{2}$ for 
$(T \equiv \beta^{-1}, \Gamma)=(0, 0.01), (0.01,0)$ (left panel), 
$(T, \Gamma)=(0,0.8),(0.8,0)$ (right panel) 
in Figure \ref{fig:fg5}. 
From these panels, we find 
that the limit cycles are getting 
collapsed as 
the strength of the noise level 
is increasing for both 
thermal and quantum-mechanical noises, 
and eventually the trajectories 
shrink to the origin $(m_{1},m_{2})=(0,0)$  
in the limit of $T, \Gamma \to \infty$.  
 \section{Summary}
In this paper, 
we considered the stochastic process 
of quantum Monte Carlo method 
applied for the quantum Hopfield model 
and investigated the quantum neuro-dynamics 
through the differential equations 
with respect to the macroscopic quantities 
such as the overlap. 
By using the present approach, one can 
evaluate the `inhomogeneous'  Markovian stochastic process 
of quantum Monte Carlo method 
(in which amplitude $\Gamma$ is time-dependent \cite{Das,Das2}) 
such as quantum annealing \cite{Kadowaki,Farhi,Morita,SuzukiOkada,Santoro,Bikas}. 
In the next step of the present study, we are planning to extend this 
formulation to the probabilistic information processing described by spin glasses 
including a peculiar type of antiferromagnet \cite{Anjan}.
\ack
We thank B.K. Chakrabarti, A.K. Chandra, 
P. Sen and S. Dasgupta for fruitful discussion. 
We also thank local organizers 
of Statphys-Kolkata VII for their warm hospitality.  
This work was financially supported by 
Grant-in-Aid for Scientific Research (C) 
of Japan Society for 
the Promotion of Science, No. 22500195. 

\end{document}